\pdfoutput=1
\RequirePackage{boolean}
\RequirePackage{reversion}

\IfFileExists
    {\jobname.cfg}
    {\input{\jobname.cfg}}
    {\input{merlin.cfg}}

\documentclass[\Anonymous{anonymous,}{}\ACM{}{nonacm,}acmsmall,10pt,screen]{acmart}
\usepackage[utf8]{inputenc}

\usepackage{authoredcomments}
\Comments{}{\UNXXX{}}

\usepackage{mathpartir}

\usepackage{pifont} 
\newcommand{\cursor}{\text{\ding{122}}}

\ACM{
\setcopyright{rightsretained}
\acmPrice{}
\acmDOI{10.1145/3236798}
\acmYear{2018}
\copyrightyear{2018}
\acmJournal{PACMPL}
\acmVolume{2}
\acmNumber{ICFP}
\acmArticle{103}
\acmMonth{9}
}{}

\ACM{
\def\trick{copyright}
\csname set\trick \endcsname{rightsretainedccbysa}
}{}

\usepackage{hyperref}
\hypersetup{colorlinks=true}

\usepackage{cleveref}

\renewcommand{\url}[1]{\textcolor{blue}{\href{#1}{#1}}}

\usepackage{listings}
\newcommand{\code}[1]{\lstinline!#1!}

\ACM{

}{}

\SillyPublisherStyle{
  \usepackage{titlecaps}

  \Addlcwords{a an the at by for in of on to up and as but or nor}

  \newcommand{\maybecap}[1]{\titlecap{#1}}
}{
  \newcommand{\maybecap}[1]{#1}
}

\newcommand{\Reason}{\href{https://reasonml.github.io/}{Reason}}
\newcommand{\Andromeda}{\href{http://www.andromeda-prover.org/}{Andromeda}}
\newcommand{\Hoogle}{\href{https://www.haskell.org/hoogle/}{Hoogle}}

\begin{document}

\title{Merlin: A Language Server for OCaml (Experience Report)}

\author{Fr{\'e}d{\'e}ric Bour}
\affiliation{}
\author{Thomas Refis}
\affiliation{
  \institution{Jane Street}
  \city{London}
  \country{UK}
}
\author{Gabriel Scherer}
\orcid{0000-0003-1758-3938}
\affiliation{
  \institution{INRIA}
  \city{Saclay}
  \country{France}
}
\email{gabriel.scherer@inria.fr}

\authorsaddresses{}

\begin{abstract}
  We report on the experience of developing Merlin, a language server
  for the OCaml programming language in development since 2013. Merlin
  is a daemon that connects to your favourite text editor and provides
  services that require a fine-grained understanding of the
  programming language syntax and static semantics: instant feedback
  on warnings and errors, autocompletion, ``type of the code under the
  cursor'', ``go to definition'', etc.

  Language servers need to handle incomplete and partially-incorrect
  programs, and try to be incremental to minimize recomputation after
  small editing actions. Merlin was built by carefully adapting the
  existing tools (the OCamllex lexer and Menhir parser generators) to
  better support incrementality, incompleteness and error
  handling. These extensions are elegant and general, as demonstrated
  by the interesting, unplanned uses that the OCaml community found
  for them. They could be adapted to other frontends -- in any
  language.

  Besides incrementality, we discuss the way Merlin communicates with
  editors, describe the design decisions that went into some demanding
  features and report on some of the non-apparent difficulties in
  building good editor support, emerging from expressive programming
  languages or frustrating tooling ecosystems.

  We expect this experience report to be of interest to authors of
  interactive language tooling for any programming language; many
  design choices may be reused, and some hard-won lessons can serve as
  warnings.
\end{abstract}

\Short{\Appendices{}{

    \titlenote{Due to space limitations, substantial portions of the
      article were omitted from this short version. We encourage the
      reader to instead consult the full article (24 pages) at
      \url{https://arxiv.org/pdf/1807.06702}.}

  }}{}

\maketitle

\section{\maybecap{Introduction}}

The research communities that study programming languages have
demonstrated their ability to propose interesting, valuable new
programming constructs and whole languages. Yet a major difficulty to
disseminate those ideas is the scarcity of tooling for those research
languages. We have built conceptual and practical tools that let us
design and implement interesting languages in a few man-years, but
building a satisfying tooling ecosystem remains much more
work-demanding, and seems reserved to industrially-backed languages.

The present report is concerned with the tooling support traditionally
provided by IDEs (Integrated Development Environments), namely
language-aware editor features, letting programmers interact with the
source of a program as they are writing, modifying or inspecting
it. Proponents of a new or niche programming language wish to provide
language support for all major IDEs and editors, and the only
practical way to do this is to build tools to share the
language-awareness logic among many different editors. We use the term
\emph{language server} to refer to a tool providing this
language-aware logic\footnote{The terminology comes from Microsoft's
  Visual Studio Code editor, which introduced a standard communication
  protocol between editors and language servers called the Language
  Server Protocol. Merlin (2013) was created before Visual Studio Code
  (2015), and does not use this specific protocol -- for now.}. Merlin
is a language server for the OCaml programming language.

From the user point of view, language servers provide instant
language-aware feedback in response to editing actions, with the
editor signaling syntax errors, typing errors and warnings, or making
auto-completion suggestions. They also enable various explicit user
actions, interrogating the partial program (what is the type of this
identifier?), navigating the project (go to definition, go to
enclosing function...), or performing editing actions (complete this
identifier, generate a covering pattern-matching on this
expression...). Adding support for a new editor to a language server
requires no language-specific logic: the editor plugin is merely
a shim that continuously sends the partial program to the server, and
exposes user-interface actions corresponding to the language-aware
server features. Merlin currently provides, through a command line
interface, language-aware editing modes for Vim, Emacs, Sublime Text,
Acme, Atom, VSCode and Intelli-J.

Let us take just one simple example, given the following, still incomplete, piece of code:
\begin{lstlisting}
let incr lst =
  List.map (fun x -> x + 1)
\end{lstlisting}
the programmer can place their editor cursor on any $x$ of the second line and
hit a key to get its type. Under the hood, the editor then invokes a command
looking something like this:
\begin{lstlisting}[language=]
ocamlmerlin server type-enclosing
  -position 2:16 -index 0 -filename test.ml -verbosity 0
\end{lstlisting}
passing the content of the buffer\footnote{A ``buffer'' is a single piece of
text or source code that is currently being edited in a text editor. Buffers
usually correspond to files on the filesystem, but they may have changed since
the last ``save'' operation or not have been written to disk at all.} on stdin.
This command returns the following JSON object on standard output, to be
interpreted and rendered by the editor appropriately
\begin{lstlisting}
{
  "class": "return",
  "value": [
    {
      "start": { "line": 2, "col": 16 },
      "end": { "line": 2, "col": 17 },
      "type": "int",
      "tail": "no"
    }, ...
  ], ...
}
\end{lstlisting}
For the experience to be smooth for the programmer, merlin, as indeed
any other language server, is expected to answer such queries in a few
hundred milliseconds, even for projects with hundreds of modules and
buffers of several thousand of lines. Besides \texttt{type-enclosing},
other examples of commands supported by merlin are
\texttt{complete-prefix} for completion, \texttt{locate} to find the
definition of a name and \texttt{errors} to display warnings and
errors.

The main technical piece required to build a language server is
a language frontend (as found in typical language implementations:
parsing and static analysis) that is reasonably incremental
(for responsiveness) and supports partial programs with missing or
erroneous program fragments. The language server continuously receives
the current partial program, and computes a partial typed-tree
-- a typing derivation for the partial program. Language-aware queries
are then implemented by processing this partial typed-tree.

One specificity of Merlin is that it was not built by implementing
a new language frontend, but by modifying existing tools to build
non-incremental frontends (for any language) to support incrementality
and partiality. These general extensions enabled many more uses than
just a language server, and we will report on some of them.

Not everything in a language server is principled; Merlin makes some
simplifying assumptions that have proven robust in practice, but also
found out that some things are more difficult or fragile than they
first seem. We hope that this experience may help other authors of
language servers.

We claim the following contributions:
\begin{itemize}
\item A presentation of how to adapt the interface of lexer and parser
  generators of the lex/yacc family to support incrementality and
  partiality, a discussion of use-cases, and a (less general)
  discussion of incremental/partial type-checking.

\item An exposition of a language server design and implementation,
  both principles and heuristics, which could be of interest to
  implementors of language servers. We present the ideas in Merlin in
  a language-agnostic way when possible, and point out when specific
  language feature or ecosystem characteristics created unexpected
  difficulties.
\end{itemize}

\SillyPublisherStyle{
  \subsubsection*{\maybecap{Article structure}}
}{
  \subsection*{\maybecap{Article structure}}
}

\begin{version}{\Short}
  \Appendices{

    Due to space limitations, we moved some of our content into
    optional appendices. We chose to focus the main article on the
    general design structure of Merlin (\Cref{sec:model}) and the
    differences compared to a traditional compiler frontend
    (\Cref{sec:frontend}), namely incrementality and partiality.

    In appendices, we discuss other adaptations to the frontend
    (\Cref{sec:frontendaux}), describe a few language-aware features
    of interest (\Cref{sec:features}), detail a few of the
    difficulties or challenges encountered during the development of
    Merlin (\Cref{sec:difficulties}), and finally provide some
    thoughts on the relations between language tooling and language
    design (\Cref{sec:thoughts}).

  }{

    We chose to focus this short version of our article on the general
    design structure of Merlin (\Cref{sec:model}) and the important
    differences compared to a traditional compiler frontend
    (\Cref{sec:frontend}), namely incrementality and partiality.

    In the full version of this article
    (\url{https://arxiv.org/pdf/1807.06702}), we also discuss the
    other adaptations to the frontend, describe a few language-aware
    features of interest, detail a few of the difficulties or
    challenges encountered during the development of Merlin, and
    finally provide some thoughts on the relations between language
    tooling and language design.

}
\end{version}

\begin{version}{\Not\Short}
  We first describe the general design structure of Merlin
  (\Cref{sec:model}) and the differences compared to a traditional
  compiler frontend (\Cref{sec:frontend}), namely incrementality and
  partiality.

  Then we discuss describe a few language-aware features of interest
  (\Cref{sec:features}), detail a few of the difficulties or
  challenges encountered during the development of Merlin
  (\Cref{sec:difficulties}), and finally provide some thoughts on the
  relations between language tooling and language design
  (\Cref{sec:thoughts}).
\end{version}

\section{\maybecap{The Merlin model}}
\label{sec:model}

In this section we will present the successive interaction models Merlin has
exposed since it was first written, these evolved over time in response to
implementation constraints and experience.

It gives a high-level perspective of how Merlin interacts with the editor, and
of the information they exchange.

\subsection{\maybecap{A reminder on the OCaml compilation model}}

OCaml source files may be either \emph{implementation} files
(\texttt{foo.ml}) or \emph{interface} files (\texttt{bar.mli}); an
implementation file contains the source of an ML module, and an
interface file contains its signature. An implementation and interface
file of the same name, \texttt{foo.ml} and \texttt{foo.mli}, together
form a \emph{compilation unit} \texttt{foo}, which is visible in the
language as a ML module \texttt{Foo} (the same name, capitalized).

When source files for a compilation unit \texttt{foo} are passed to
the compiler, it produces some compiled implementation files, and
a compiled interface file \texttt{foo.cmi}; if the compiler is given
a file which uses the module \texttt{Foo}, it only needs to access
\texttt{foo.cmi} to type-check this use. This separation of
implementation and interface enables incremental and separate
compilation.

Compiling the source of a compilation unit requires knowing about the
other units that it depends on. The compiler runs in a given
compilation environment, which is a mapping from module names to
compiled units on the filesystem. (In particular, this compilation
environment determines the initial typing environment to type-check
the current compilation unit.) This mapping is built from the
already-compiled files in the current project and from the installed
third-party libraries whose path was passed to the compiler at
invocation time.

OCaml enforces an acyclic dependency graph between compilation units:
a program or library is an ordered list of units, each depending only
on units before it -- and, in the case of libraries, on third-party
units to be linked separately.

\subsection{\maybecap{The old build and the buffer}}

When the user is editing a source file buffer with
a Merlin-enabled editor, Merlin will continuously receive the partial
buffer state from the editor. Just like the compiler, it reconstructs
the compilation environment from compiled interface files, found
either in the current project (typically coming from an old build of
the project) or from third-party libraries.

From the point of view of Merlin, there is thus a single buffer being
edited, which requires special incrementality and partiality support,
and a set of compiled interface files that are used in the standard
way to build a typing environment.

Merlin supports editing several file buffers at once, but each buffer
belongs to an independent \emph{session}, with no cross-session
communication.  Merlin never tries to synchronize or share partial
information from several files being edited simultaneously. In
practice this heuristic simplification works very well, for at least
two reasons. First, each session typically has access to
a mostly-correct interface for the other files coming from an old
build. Second, programmers typically spend time editing implementation
files \texttt{foo.ml} without breaking the interface file
\texttt{foo.mli}, allowing new builds to create correct
a \texttt{foo.cmi} even if the whole build fails on partial files.

\subsection{\maybecap{Evolving the interaction models}}

\subsubsection{\maybecap{The toplevel model and linearization}}
\label{subsub:model-toplevel}

The first interaction model used by Merlin was inspired by interactive
toplevels (read-eval-print loops). Merlin would interpret sentences of
the buffer in order, up to the first point with a hole or an
error. Early version would even show a ``blue zone'', inspired by the
\Xgabriel[Proof General]{cite the proof general paper?} interface, to
provide visual feedback on which prefix of the buffer had been
type-checked. A first prototype of such an approach is easy to build
on top of a stock parser found in an interactive toplevel -- the OCaml
compiler distribution comes with such a toplevel, who shares its
grammar description with the compiler. This model also offers some
weak error-resilience (partiality support) for cheap: phrases
containing an error can simply be skipped.

The main difficulty with this model is that the granularity of
``toplevel phrases'' used by toplevels is much too coarse to support
partial programs in practice. In ML languages with a hierarchical
module system, a phrase may be a complete module declaration with
several sub-phrases. Consider the following example of partial program
(at least an \code{end} terminator for \code{struct .. end} is missing):
\begin{lstlisting}
module IntOrd = struct
  type t = int
  let compare m n =
    if m < n then Lt
    else if m > n then Gt
    else Eq
\end{lstlisting}
This incomplete module declaration is an incomplete/invalid toplevel
phrase, but a user would still expect the tool to process the
declarations of \code{t} and \code{compare}.

To satisfy this need, Merlin moved to a ``linearized'' grammar for
OCaml, where nesting structures are turned into independent sentences
allowing partial processing. A single-phrase module declaration
(here we use the ugly optional OCaml phrase terminator
\texttt{;;} for clarity)
\begin{lstlisting}
module M = struct
  phrase_1;;
  phrase_2;;
end;;
\end{lstlisting}
is represented, after linearization, by a series of phrases
\begin{lstlisting}
/(enter_module_struct)/ M;;
phrase_1;;
phrase_2;;
/(leave_module)/;;
\end{lstlisting}
where \code{/(enter_module_struct)/} and \code{/(leave_module)/} are
new Merlin-specific phrase-forming constructs that linearize the
\code{module M = struct ... end} nesting -- this is similar to the
transformation of a tree grammar into a streaming interface. Note that
this need for linearization is not specific to nested ML modules, it
naturally arises with all nesting structures such as local
declarations or list or record literals.

(This form of linearization is also present in Proof General
interfaces, as we often want to have the ``blue/checked zone'' end in
the middle of a nesting construct. The \code{/(Proof)/} and
\code{/(Qed)/}/\code{/(Defined)/} commands of Coq are precisely
linearizations of a document nesting structure.)

The Merlin authors modified the upstream toplevel grammar to parse
this linearized grammar. However, they quickly found out that the
effort to linearize nesting constructs was tedious, incomplete, and
consuming more and more time. Not only must the parser be modified,
invasive changes were also required in the type-checker to deal with
these new constructs, with a different approach required for different
kind of linearizations. The development and maintenance burden was
much too important, and a new approach was necessary.

\subsubsection{\maybecap{The partial compiler model}}
\label{subsub:model-partial-compiler}

The next iteration of incrementality support for Merlin used
a standard LR grammar description for the OCaml grammar, with a parser
generator modified to generate an incremental parser -- we describe
this design in \Cref{subsec:incremental-parsing}.

This changed the interaction model of Merlin from acting as an
(type-checking-only) interactive toplevel to acting like
a (type-checking-only) compiler: the editor passes source buffers to
Merlin, along with information on the compilation environment, and
Merlin does its best to produce partial typing information from
a partial Abstract Syntax Tree (AST) -- stopping at the point where
the incomplete buffer ends, where the first error was encountered, or
where the user cursor was editing, whichever comes first.

Note that it does not matter, conceptually, whether such a ``partial
compiler'' is fed incremental or non-incremental output from the
editor: the whole pipeline is incremental and can recompute a result
from previous states and a source change, but this source change may
either be communicated by the editor itself or computed by Merlin from
the entire new buffer (copying source text in memory is not
a performance bottleneck) and previous states.\footnote{How Merlin
  records previous states is an implementation detail. Merlin works as
  a daemon, so it keeps a cache in memory, but a short-running tool
  could persist state on the disk. The important property to ensure is
  that the presence or absence of saved state does not modify
  Merlin's results, only its performance. In fact, Merlin does provide
  a short-running mode that does not use any cache, which is
  slower but helps testing correctness.}

The first iteration of this ``partial compiler'' represented partial
programs with parser states, which essentially corresponds to an AST
spine with the right-most nodes missing. Merlin implemented a variant
of the type-checker that works on these partial ASTs, rather than
complete ASTs as in a compiler -- we detail this in
\Short{
  \Appendices{\Cref{subapp:incremental-typing}}{the long version of this article}
}{\Cref{subsec:incremental-typing}}.
This requires a bit of work
for each AST construction, but adapting the type-checker code was
surprisingly easy.

On the other hand, the logic to traverse the parser state and expose
the partial AST to the type-checker proved cumbersome to synchronize
with changes in the continuously-evolving OCaml compiler fronted. This
approach would be a sensible choice if this piece of infrastructure
was included in the upstream language implementation, but Merlin
needed a more pragmatic approach.

\subsubsection{\maybecap{The completed compiler model}}
\label{subsub:model-completed-compiler}

Merlin's technology for incremental parsing stabilized relatively
quickly; an aspect of the parsing stack that improved gradually over
time and keeps evolving today is error recovery, and in particular
\emph{completion}, which consists in completing the valid prefix of
a buffer into a more complete program, either by just inventing new
tokens or non-terminals to close off nesting structures -- when
reaching the end of the buffer -- or by also trying to skip some parts
of the buffer and reconnect later, after the error error or hole. We
detail completion mechanisms in \Cref{sec:recovery}.

Once syntax completion got good enough to work reliably in practice,
it became possible to abstract away the partial nature of the input
AST: a partial syntax tree is completed by recovery. Instead of having
the type-checker work on partial ASTs, it is thus possible to reuse
a normal type-checker expecting ASTs. Adaptations to the type-checker
are still necessary for error-resilience (some parts of the parsed
syntax may not type-check, and the completed parts of the syntax may
also break type-checking locally), but the input format is the same as
for the standard compiler, and this facilitates code reuse.

\subsubsection{\maybecap{Memoization in absence of incrementality}}

Recent versions of Merlin have tried to improve support for
pre-processing mechanisms used in the OCaml community -- the
\texttt{ppx} syntax extensions. Those extensions rely on running
external programs to transform syntax trees between the parsing and
the type-checking phase; the API (Application Programming Interface)
of these AST-rewriters is still evolving as tooling changes, but they
currently do not provide an incremental interface. A preprocessor is
fed the entire AST to transform, and may generate different code at
the beginning of the tree depending on what happens later in the code,
so the entire transformation result must be recomputed on each source
change.

Discussion is ongoing with designers of syntax preprocessing libraries
to enable incrementality, by letting extensions produce
a serialization of their internal state after processing a part of the
buffer, and advertise that their transformation commutes with addition
of further phrases of input.

In the meantime, Merlin behaves in a non-incremental way on files
which use syntactic preprocessing, using memoization to avoid as much
recomputation as possible during the type-checking phase. Merlin
considers the post-processed source a sequence of toplevel phrases,
and saves the typing environment for each prefix sequence of
phrases. After a buffer change, preprocessors may perform arbitrary
change to the syntax tree, but if the result is close to the previous
postprocessed output Merlin will only re-type-check the phrase
currently being edited.

\subsection{\maybecap{Remark: purity prevails again}}

One notable design and implementation consequence of the move from
a toplevel-like model to a compiler-like model is purity.

In the toplevel model, phrases are naturally understood as stateful
actions on the typing state -- linearization turns declarative
nesting structures into more stateful commands. Each Merlin session
would keep the ``toplevel state'' of the buffer on its end, and
various bugs showed up where the synchronization between editor state
and Merlin state would break. Typically, knowing what to replay after
a Merlin crash would require delicate and buggy communication with the
editor.

In the compiler models, the interface between the editor mode and
Merlin is much simpler: the editor sends the text buffer to Merlin, as
one would invoke the compiler on the file. Merlin's behavior is a pure
function of its input; an internal cache is used to enable
incrementality, but its presence only affects reactivity, not the
output content, so a desynchronization at worst results in a slowdown.

We strongly recommend that language servers follow this pure
model. The improvements in consistency for users and debuggability and
maintenance for developers are considerable. This interface even
allows hot-reloading a new Merlin version during an editing session,
which proved very useful to develop Merlin itself.

\subsection{\maybecap{The language and the model}}

\begin{figure}[!htb]
  \newcommand{\lines}[1]{\begin{tabular}{c} #1 \end{tabular}}
  \begin{tabular}{|c|c|c|}
    \hline
    & \textbf{definition before use} & \textbf{forward references}
    \\
    \hline
    \textbf{definitions at toplevel}
    & \lines{easy, linearized prefixes
             \\ (C)}
                                   & \lines{medium, full linearized
                                            \\ (Haskell, Erlang, Java) }
    \\
    \hline
    \textbf{rich scoping structures}
    & \lines{medium, fine-grained prefixes
            \\ (ML, Lisp)}
                                    & \lines{hard, full fine-grained
                                            \\ (Scala)}
    \\
    \hline
 \end{tabular}
 \caption{Language design aspects and incrementality model}
 \label{fig:languages-and-models}
\end{figure}

In all these interaction models that Merlin went through, Merlin works
best on the prefix of the buffer before the first error or
incompleteness. Completion and recovery strategies help analyzing
further than the first error, but there are always difficult cases
where this situation is degraded. Furthermore, checking only up to the
cursor, the place where operations and queries happen, is
a robust simplification to speed up processing.

The fact that this strategy works well in practice relies on
a particularity of the ML-family programming language: it is
relatively rare for identifiers to be used before their
declaration. ML languages do contain mutually-recursive declarations
that allow forms of forward references, but user programs tend to
make a moderate, local use of them.

In contrast, many languages, typically Haskell, Erlang or some
object-oriented languages, rely on liberal uses of recursion nests;
the model of working only with definitions above in the buffer does
not lead to an acceptable user experience. On the other hand,
linearization works much better in these languages, that have a simple
top-level structure with less nesting of module/scoping constructs --
one may linearize the whole buffer, dropping incomplete/erroneous
phrases. Languages with both rich scoping structures and abundant
forward references, like Scala, would be harder to cope with. We
summarize this design space in
Figure~\ref{fig:languages-and-models}.

\section{\maybecap{Changing the frontend}}
\label{sec:frontend}

The Merlin authors explicitly tried not to reimplement a new OCaml
frontend, but to reuse existing tools and adapt them to support
incrementality -- and recovery. The specific tools used are
implemented in OCaml and used within the OCaml community, but they
follow a common heritage of \texttt{lex}-inspired lexer generators and
\texttt{yacc}-inspired parser generators -- and an extended
implementation of the Hindley-Milner type inference algorithm. There
are similar tools within most language communities.

\Short{}{
\subsection{\maybecap{Incremental lexing: adapting OCamllex}}
\label{subsec:incremental-lexing}
}

\newcommand{\SectionIncrementalLexing}{\begingroup
The OCaml compiler uses a lexer defined using the OCamllex
tool, which was not designed to allow incremental lexing. Merlin
authors found that a minimal extension to OCamllex allowed
them to reuse the compiler's lexer definition.

OCamllex takes as input a series of rules, mapping regular
expressions to lexical tokens:
\lstinputlisting{code/incremental_lexer_1.ml}

It compiles each rule into a token-producing regular automaton,
exported as a function of type \code{Lexing.lexbuf -> user_token}
where \code{user_token} is whatever token type the user's lexer
returns. \code{Lexing.lexbuf} is a standard type describing the lexing
input buffer, that can be produced from a file handle, a pipe or an
in-memory string by the \code{Lexing} library. The compiled lexing
function imperatively updates the lexical buffer, calling I/O
functions to fill an input buffer if necessary, and advancing position
marks so that the function can be called in a loop on a given lexing
buffer to produce a stream of tokens.

To support incremental and partial lexing, a lexer must be turned into
a pure function. The behavior of the generated lexers was already
mostly pure, in that it does not depend on global state outside the
\code{lexbuf} parameter: if the lexer is invoked several times with
a \code{lexbuf} in the same state, it produces the same
result. However, calling the lexing function may implicitly perform
I/O actions (consuming inputs from a file, etc.) when the lexing
buffer is empty and needs to be refilled; besides, if no more input is
available at this point, the lexing function fails with an
exception. A refill may be required in the middle of lexing a long
lexeme (typically comments and string literals), so the lexing state
on failure may have been mutated before the failure.

The Merlin authors introduced \emph{refill handlers} to let lexer
authors specify alternative ways to handle buffer-refill
actions. A handler can be specified by a \texttt{refill} directive in
the lexer description
\lstinputlisting{code/incremental_lexer_2.ml}
which specifies what to do when a refill is necessary, taking the
lexing buffer and a continuation of the lexer's action. The default
handler just calls the lexing buffer's effectful refill function, and
then calls \code{k lexbuf} to continue. (The generated code assumes, for
performance, that the continuation \code{k} is called at most once.)

This corresponds to an effect handler for an effectful \code{Refill}
operation. In particular, one can write lexers parametrized over
an arbitrary monad (notice how the OCamllex-generated code is
placed inside the body of an ML functor):
\lstinputlisting{code/incremental_lexer_3.ml}

\subsubsection{\maybecap{Integration and other uses}}

Refill handlers are a non-invasive change, adding 82 lines to the
OCamllex codebase. After some incubation time within the Merlin
codebase, they were submitted for upstream inclusion in February 2014,
and were included in OCaml 4.02 released in August
2014.\footnote{\url{https://github.com/ocaml/ocaml/pull/7}.} Their
main use-case outside Merlin is to build lexers that work well with
aysnchronous I/O libraries, in particular using the Lwt cooperative
concurrency monad.

\paragraph{Remark:} If you had to implement a Merlin-like tool without
reusing an existing lexer generator -- or you were designing a new
lexer generator -- we would rather recommend trying to produce pure
lexers where the users are directly in control of refilling actions,
which corresponds to the free \code{\{Return, Refill, Fail\}}
signature. The user then has to implement the event loop that receives
\code{Refill} and \code{Fail} actions and acts accordingly. This loop,
frequently transferring control from the lexer back to the user, adds
some performance overhead, but we believe that a careful
implementation can be competitive with current interfaces where the
lexer is always in control.
\endgroup}

\Short{}{\SectionIncrementalLexing}

\subsection{\maybecap{Incremental parsing: adapting Menhir}}
\label{subsec:incremental-parsing}

The first toplevel-like model for Merlin
(see \Cref{subsub:model-toplevel}) used a very simple incremental
parsing strategy: it would parse the buffer phrase by phrase -- using
existing support in the OCaml grammar for toplevel phrases -- and
catch the exceptions raised on reaching end-of-input. After an editing
action, only the edited phrase needed to be reparsed. When Merlin
moved to a compiler-like approach (\Cref{subsub:model-partial-compiler}),
a new approach to incremental parsing was required.

The OCaml compiler uses the OCamlyacc parser generator, inspired from
its C counterpart Yacc. It takes a grammar description in Yacc style
(including a list of tokens), and generates a parser for each declared
start symbol.

\begin{lstlisting}
{ (* optional prologue *)
type tree =
  | Leaf of int
  | Plus of tree * tree    }
$\%$/(token)/ INT<int>
$\%$/(token)/ PLUS
expr:
  | n=INT { Leaf n }
  | a=expr PLUS b=expr { Plus (a, b) }
\end{lstlisting}

The generated parser for this grammar will have the following interface:
\begin{lstlisting}
exception Failure of Lexing.position
val expr: (Lexing.lexbuf -> token) -> Lexing.lexbuf -> tree
\end{lstlisting}
The generated parser takes a (\texttt{lexbuf}-consuming) lexer and
produces a \texttt{lexbuf}-consuming parser. This parser will
repeatedly invoke the lexer on the input buffer, producing a complete
AST tree out of the resulting token stream -- or raise an exception on
parse error.

As for lexers (%
\Short{
  \Appendices
  {\Cref{subapp:incremental-lexing}}
  {see the long version of this paper}
}{\Cref{subsec:incremental-lexing}}
), the generated
function assumes full control over the parsing process\footnote{The
  in-control interfaces for lexing and parsing generators come from
  the original tools for C -- where the lack of algebraic datatype
  makes finer-grained control cumbersome -- and also from performance
  concerns: giving full control to the generated consumer lets it
  easily use local mutable state without paying serialization
  costs.}, and we need to give control back to the user to bring
support for incrementality and partiality. The Merlin authors
developed an incremental parsing interface for Menhir,
a parsing generator that is compatible with OCamlyacc's
grammars; it is implemented in OCaml rather than C, so more convenient
to develop and extend. They proposed a new ``incremental mode'' for
Menhir, that generates a different interface:\footnote{We
  simplified the interface for the presentation: there are several
  \texttt{Intermediate} states, see the
  \href{http://gallium.inria.fr/~fpottier/menhir/manual.pdf}{Menhir
    manual} (Section 9.2, Incremental API).}
\begin{lstlisting}
(* generic incremental interface *)
type 'a env (* abstract parser state *)
type 'a checkpoint =             (* intermediate results: *)
  | Result of 'a                 (* successful parse *)
  | Error                        (* parse error *)
  | Intermediate of 'a env       (* more parsing actions required *)
(* the generated transition function *)
val step : 'a env -> token -> 'a checkpoint
(* grammar entry point(s) *)
val expr_entry : tree env
\end{lstlisting}

Most Yacc-style parser generators, Menhir included, use a parsing
strategy in the LR family (LR, LALR, etc.). They generate a parsing
automaton (a deterministic pushdown automaton). The parsing state is
represented by a state number and a parsing stack (the innermost
construction being parsed is at the top). Transitions (the dreaded
\texttt{shift} and \texttt{reduce} transitions) go from one state to
another, some requiring to consume a token from the input. In this
incremental interface, \code{'a env} is an abstract type for a parsing
state that will eventually return a semantic value of type \code{'a}
or fail with an error, and the \code{step} function consumes a token
and performs all transitions until a new token is required.

With this incremental interface, the user is in control of the
parsing, feeding tokens one after another. The original interface is
easily expressible as a parsing loop:

\begin{lstlisting}
let expr token buf =
  let rec loop = function
    | Result x -> x
    | Error -> raise (Failure (Lexing.curr_pos lexbuf))
    | Intermediate env -> loop (step env (token buf))
  in loop expr_entry
\end{lstlisting}

States and checkpoints are purely functional: they can be stored and
reused at will. In particular, we get incrementality by caching them,
indexed by token location: when a part of the buffer is modified, we
can restart from the last checkpoint before it. There is a lot of
sharing within parsing stacks and the semantic values on the stack,
keeping the memory usage close to non-incremental parsing. We are
assuming, of course, that semantic actions -- the user-provided code
running to build AST values -- are pure as well, as they should be.

\subsubsection{\maybecap{Inspecting the parsing stack}}

With the new interface, the state of the parser is available as
a value. Merlin authors developed an advanced API to inspect
(and modify) this value outside of the normal parsing process.

Giving access to the state of the parsing automaton, which is
represented as an integer, is easy, but giving type-safe access to the
semantic values on the parsing stack requires care. Each element of
the parsing stack corresponds to a symbol (a terminal or
non-terminal symbol), and contains the corresponding semantic value:
\begin{itemize}
\item for terminals, it is an arbitrary value passed by the lexer to
  the parser (e.g. when parsing an integer, the INT token will be
  generated together with the value that was parsed)

\item for non-terminals, it is the value returned by the semantic
  action of a production rule (for this non-terminal) that matched
  the elements already present on the stack.
\end{itemize}

In particular, semantic values of different stack items may be at
different types -- in our example, \code{int} and \code{tree}.

\paragraph{Remark:} This problem was discussed in detail precisely in
the original article presenting Menhir,
\citet*{pottier-regis-gianas-typed-lr}, which remarked that
GADTs (Generalized Algebraic Datatypes) could be used to give
a fully-typed presentation of an LR parser stack. At the time, OCaml
did not have GADTs -- they became available in OCaml 4.00, released in
2012 -- so Menhir internally used unsafe type casts.

The Merlin authors extended Menhir to generate a stack-inspection
interface using GADTs to type-check semantic values in a precise
way. For our example, the following additional code is generated:
\begin{lstlisting}
type 'a terminal =
  | T_INT : int terminal
  | T_PLUS : unit terminal
type 'a nonterminal =
  | N_expr : tree non_terminal
type 'a symbol =
  | T : 'a terminal -> 'a symbol
  | N : 'a nonterminal -> 'a symbol
type element = Element : 'a symbol * 'a -> element
val pop : 'a env -> 'a env option
val top : _ env -> element
\end{lstlisting}

The \code{'a terminal} and \code{'a nonterminal} types contain one
constructor per symbol, but they do not carry any data; they merely
constrain their type parameter \code{'a} to be equal to the type of
the corresponding semantic value. The \code{element} type packs an
\code{'a symbol} together with the semantic value \code{'a}, for an
existential variable \code{'a} -- this is a typical GADT encoding of
a strong dependent pair, here
$\Sigma(s : \mathsf{Symbol}).\,\mathsf{SemType}(s)$. The function
\code{top} provides access to the symbol and semantic value at the top
of the stack, and \code{pop} removes one element off the stack.

\paragraph{Remark:}
The GADT proposed in \citet{pottier-regis-gianas-typed-lr} contain
more fine-grained information than the types above. Not only do they
allow to distinguish each symbol of the grammar, they include
different constructors for each state in the parsing automaton,
which keeps fine-grained typing information on the transitions from
one state to another. In particular, it lets us statically
predict, from the element on the top of the stack, which elements
may be below it on the stack.

The simpler GADTs used by Merlin does not suffice to give
a fully-typed presentation of the stack: \code{'a env} remains an
abstract type with more information, and \code{top} is a lossy
projection. On the other hand, they are solely determined by the
grammar structure (which the user wrote) rather than the parsing
automaton (produced by the compiler), so they are much simpler to
understand and use.

\subsubsection{\maybecap{Integration and other uses}}

The work on an incremental Menhir interface started within Merlin in
December 2013. It was adapted by Fran{\c{c}}ois Pottier and merged in the
upstream Menhir code in December 2014. Since its integration, three
major use-cases were:
\begin{enumerate}
\item Error message support. The complete API for the parser state
  exposes the automaton state number, which makes it easy to index an
  error message on the error state -- an approach first proposed by
  \citet*{jeffery}. More complete approaches are possible by
  inspecting the parsing stack; in particular, it enables fetching AST
  fragments around the error to print their location or pretty-print
  their source code in the error message. This is demonstrated in
  \citet*{pottier-reachability-cc-2016}, which builds on Menhir's
  incrementality support and typed stack API.

  Other error-message strategies were implemented on top of the stack
  introspection APIs; we know of developments within Merlin itself,
  and in the frontend of the \Reason{} language at Facebook.
\item Generalized lexer hacks. Some languages (C and Bash are
  notable example) are specified with a separate lexer and parser, but
  actually require a communication from the parser to the lexer, as
  the same source text may need to be recognized as different tokens
  depending on the current parsing state -- this is called the ``lexer
  hack'', traditionally implemented through subtle use of mutable
  state. Using Menhir's incremental parsing API, one can pass the
  parsing state as a lexer parameter, and then use parsing stack
  inspection in the lexer as a pure approach to parser-lexer
  communication. This is used in the Bash parser of
  \citet*{bash-parser}.

\item Simulating local GLR parsing. GLR parsing can parse
  arbitrary context-free grammars by forking the parsing along
  different branches on ambiguities. The pure parsing interface allows
  to simulate this by doing the backtracking by hand, feeding the same
  parsing state to different continuations. This is less efficient
  than GLR algorithms that share more data between parsing threads, but
  it is very convenient to deal with local/bounded non-LR
  ambiguities. This is used in the \Reason{} codebase, and also in
  \citet*{bash-parser}.
\end{enumerate}

\Short{}{
\subsection{\maybecap{Incremental typing, sort of}}
\label{subsec:incremental-typing}
}

\newcommand{\SectionIncrementalTyping}[0]{\begingroup
While the OCaml typechecker was not designed for interactive use as
done by Merlin, the interface required by the OCaml toplevel provided
a reasonable starting point.

The typechecker is a mixture of pure and impure code, with unification
and external module loading being the notable impure features.
However, the toplevel needs to be able to rollback changes when a
sentence fails in the middle of typechecking. The typechecker is thus
augmented with a snapshot feature which is able to undo changes
done by unification if necessary.

\lstinputlisting{code/incremental_typing_1.ml}

This snapshot feature proved enough for most needs: after a buffer
change, Merlin restarts typing from the last snapshot before the
change. Under the hypothesis that edits happen next to each other,
this gives a small bounded amount of typing work on each edit.
\endgroup}

\Short{}{\SectionIncrementalTyping}

\subsection{\maybecap{Partial parsing with error recovery}}
\label{sec:recovery}

Recovery is the ability to handle partially incorrect files in Merlin,
by asking the parsing and typechecking code to recover from unexpected /
erroneous or unterminated input. Error resilient parsing is done by
meta-programming the parser, using an extension to Menhir. On the type
checking side, the changes are more mundane: the ability to recover
from type errors in a few select points as well as a few heuristics to
get a useful type derivation from the correct parts of the program.

From a software design point of view, recovery changes the possible
outcome of the frontend from either a typed AST or an exception
(representing a lexing/parsing/typing error) to a typed AST \emph{and}
a list of exceptions. This requires Merlin to produce an AST for any
input and to type this AST, however wrong it is. In exchange, the rest
of Merlin features always expect a typed AST to be available, although
it may contain incorrect sub-derivations.

The parser enters a recovery state when receiving a token that is not
expected in the current context.
\Xgabriel{TODO gasche: explain, fred: is it necessary? it is the "viable prefix
property", do we need to remind the reader?}
LR parsing guarantees that an error is detected as early as possible: the
erroneous token is the first one for which there is no valid parse. The
recovery algorithm makes use of this property by not trying to backtrack in
general: all the input that has been parsed is kept -- even though the most
satisfying correction for the user could have been to drop
a token. The recovery will only try to complete the parse by filling
holes, either by synthesizing values or by resuming the parse using
tokens coming later.

The details of how the Merlin authors guide OCaml parsing error
recovery, described in the following sections, rely on both
language-agnostic algorithms and heuristics that are guided by
annotations on the language grammar. This approach should work in any
language server, and in particular it is easy to add recovery
capabilities to a parser, for any language, written using the Menhir
parser generator.

\Xgabriel{fred: en fait reason utilise la recovery yacc pour l'instant ;
des gens ont déjà demandé pourquoi c'était moins robuste qu'avec OCaml,
mais on a rien fait pour l'instant}
In particular, the Merlin authors are adapting this work to improve the
frontend of the \Reason{} language -- which is supported by Merlin --
and studied the needs of the prototype \Andromeda{} proof assistant
being designed in Ljubljana -- which hasn't adopted the Merlin tooling
yet.

\Short{}{
\subsection{\maybecap{No need for lexing recovery}}
\label{subsec:lexing-recovery}
}

\newcommand{\SectionRecoveryLexing}{\begingroup
All steps of the pipeline should be able to process incorrect
input. However, the logic of the lexer being very simple, Merlin does
not try hard to handle failures from the lexer. The error is logged,
and the lexer is resumed on the remaining characters, until a token
can be lexed or end-of-file is reached.
\endgroup}

\Short{}{\SectionRecoveryLexing}

\subsection{\maybecap{Parsing recovery: synthesizing values}}

Consider the case where no more input is available (for instance, the
user is typing a new file and only a prefix is available). The current
parse represents an AST being filled from left to right, where all
nodes on the right are missing at some point.

The authors of Merlin wrote a Menhir extension that automatically
completes this partial AST, guided by the grammar and a few extra
annotations. The technical details are out of the scope of this
article, but the main idea is that, from a parser continuation, it is
possible to synthesize a new symbol (terminal or non-terminal) to
continue parsing, such that this repeated process always produces
a complete AST after a finite number of steps.

In the general case, many different completions are possible, and some
will produce completed parse trees that are more likely to make the
user-written segments well-typed -- and thus enable type-level
feedback on user code -- than others. To guide the recovery process,
the Merlin authors annotated the input grammar to specify a ``cost''
for each shift/reduce transition happening during completion --
transitions that have lower chances of producing a sensible program
are given a higher cost. The completion algorithm then searches for
a completion sequence of minimum combined cost.

\subsubsection{\maybecap{Recovery annotations}}

To be able to produce an AST from a completed parse tree, each
synthesized symbol must also come with a semantic value; this is also
provided by annotations. For example, Merlin's OCaml grammar contains
the following annotated definitions
\begin{lstlisting}
$\%$/(token/) <string> STRING [@cost 1] [@recovery ""]
match_cases [@recovery []]:
    match_case { [$\$$1] }
  | match_cases BAR match_case { $\$$3 :: $\$$1 }
\end{lstlisting}

The annotation \code{[@cost 1]} gives a relatively low cost to
synthesizing a missing STRING token, and \code{[@recovery ""]}
suggests pretending that the empty string was parsed. The
\code{match_cases} nonterminal is also provided with a semantic value
to recover (the empty list of clauses). Synthesizing a complete
non-terminal at once is much more robust than trying to complete with
tokens only.

Semantic actions must be well-behaved for the recovery to work properly:
\begin{itemize}
\item Since the recovery can explore different branches and thus
  execute user code multiple times, actions should be observably pure.
\item Errors should be logged and not affect the parser control
  flow. If an action can fail (raise an exception), it should be
  annotated with an infinite cost to avoid selecting it during
  recovery.
\end{itemize}

\subsubsection{\maybecap{Integration and Adoption}}

The generic annotation mechanism, and the ability to postprocess them
in tandem with a description of the generated parser automaton, were
developed for Merlin's fork of Menhir over the course of 2015, and
upstream in 2017. We are not currently aware of adoption by others
than Merlin authors, but the Menhir designer, Fran{\c{c}}ois Pottier,
seems relieved that some of the features requested for Menhir can now
be implemented as third-party annotation processors.

\subsection{\maybecap{Parsing recovery: reconnecting to user input}}

After encountering errors in the input file, the completion mechanism
will synthesize symbols, but it would be nice to resume using the
normal input as soon as possible. There is a tension between filling
the AST, which will succeed but may drop the user input after the
error, and resuming parsing, which is less reliable but will offer
a better user experience if successful.

At any point, the recoverer has the choice between synthesizing one
more value or consuming a token. Systematically trying both would lead
to unacceptable performance. The main heuristic that drives Merlin's
implementation is indentation, on the assumption that lexical
indentation follows nesting of constructions in the grammar.

Precisely, Merlin compares the column of the current token and the
column of the item at the top of the parser stack:
\begin{itemize}
\item If the parser is more indented, Merlin considers that the token
  should attach to a construction that is outside of the one the
  parser is currently parsing, so Merlin will do a synthesis step.
\item If the parser is at a lesser or equal level of indentation,
  Merlin will try to consume the token, and if this fails, drop the
  token and fallback to synthesis.
\end{itemize}
While the indentation assumption is not true in practice for all
indentation styles, this heuristic gives a satisfying behavior most of
the time and should work for many languages.

\Short{}{
\subsection{\maybecap{Typing recovery}}
\label{subsec:typing-recovery}
}

\newcommand{\SectionRecoveryTyping}{\begingroup
The typechecker of OCaml is a large piece of code that is actively
developed independently of Merlin. It is important for Merlin to
match its behavior closely.

Merlin contains a copy of the typechecker of each version of OCaml
that is supported, with a small list of patches on top. There are two
sorts of patches:
\begin{itemize}
\item critical and generic changes to synchronize the state of
  Merlin and the typechecker, and
\item finer-grained changes to enable more features or improve the
  precision of Merlin answers.
\end{itemize}

When a new version of OCaml is made available, the critical patches
are ported. The other changes come later to improve the
experience. This enables using Merlin on the development version of
the compiler with a small amount of work.

\subsubsection{\maybecap{Critical typer changes}}

The first required change to the type-checker is a finer-grained
mechanism to keep track of assumptions made on the filesystem
state. The OCaml type-checker will cache compiled interface files
accessed during the type-checking process. Merlin is a long-running
process, and users rightfully expect it to react to recompilations
that change these files on the disk. Merlin regularly checks whether
a cached compiled interface changed on the filesystem, or if a new
file was added that could shadow the name of an existing module, and
recreates a coherent typing environment.

The second change is a move from a first-failure report to reporting
all type errors. The core of the typechecker is implemented by a few
recursive functions to type expressions, patterns and modules.  Merlin
sneaks into the recursion to catch calls that raises an exception: log
the error, and resume the typechecking. Thus, if, for instance,
a unification failure causes a part of an expression to fail to
typecheck, the corresponding partial derivation tree is incorrect, but
the rest of the expression can be typechecked.

Finally, the third changes pertains to creating typed AST nodes even
for ill-typed code. If a subterm is ill-typed, dropping the whole
typing sub-derivation would provide a bad experience, as it makes the
corresponding AST nodes, and in particular the source positions stored
on the nodes, absent from the returned typed AST. Interactions from
the editor will often refer to a source position in the buffer, and
for the interaction to succeed this position must be matched with
a node in the typed AST. It is better to keep ill-typed nodes in the
typed AST, even if the types are unsound.

When type-checking a subterm fails, Merlin then generates a ``fake''
node in the typed AST, whose typing judgment is exactly the one
expected by its context, and whose children premises are the list of
partially typed AST fragments that were type-checked for this subterm
before the failure occurred. This fake node preserves those typed
fragments whose connection to the rest of the type-tree is unknown.

\subsubsection{\maybecap{Fine-grained changes}}

The generic rules above do the bulk of the work, however not all OCaml
construction can be handled in this way.

For instance, the typechecking of a record expression depends both on
the lookup of field labels and the typechecking of subexpressions. If
the typechecking of labels fail, the typechecker gives up early and
never produces the sub-derivations for each expression. Merlin
instruments the typechecker to consider that a missing label is not
a fatal error, but rather continue under the assumption that this
label should be added later for the program to be correct. Merlin does
other similar changes on a case-by-case basis, according to practical
relevance and development costs.

This change of perspective from ``is this subterm correct with
according to my environment'' from ``would this subterm be correct in
a future extension of my typing environment?'' could benefit from more
theoretical research, and seems related to the research arc on
editor-inspired type systems proposed by \citet*{cyrus-omar}.
\endgroup}

\Short{}{\SectionRecoveryTyping}

\Short{}{
\section{\maybecap{Features of interest}}
\label{sec:features}
}

\newcommand{\SectionFeatures}{\begingroup
\subsection{\maybecap{What is the type at the cursor?}}
\label{subsec:type-at-point}

This feature is surprisingly delicate.

First, it is not necessarily clear which language object the user is
pointing at. In the source expression \code{List.length}, the user may
want the type of the \code{length} function, but also sometimes the
signature of the \code{List} module. Merlin will print the latter if
the cursor is before the dot.

Decomposing the structure of names in this way required to refine the
source position information stored by the OCaml AST for complex
names -- in the example above, the \code{List} component must have its
own location for the position-based decision of what is under the
cursor to work.

Second, in presence of overlapping namespaces, the same identifier may
be one of many sorts of name, including a type name, an identifier
name or a record field label. Merlin must use the local parsing
context to disambiguate the syntactic category under the cursor --
this is easily done using the API for parsing state introspection
described in \Cref{subsec:incremental-parsing}, popping nodes
off the LR stack until we find one of the non-terminals unambiguously
identifying a specific namespace.

On the other hand, sometimes using the local context is the wrong
choice: if the current phrase contains a type error early on, the
type-checker may not have been able to type-check the expression under
the cursor, and would return an unsatisfying inference variable. In
this case, Merlin also provides the type that the expression would
have in the global environment of the compilation unit. Providing the
global type may also return a more polymorphic type: an occurrence of
the empty list \code{[]} in a specific environment may have an
instantiated type such as \code{int list}, while the global type is
the more informative \code{'a list}.

In presence of type aliases, there may be several valid equivalent
types for a given expression. Merlin's default behavior is to show the
same type than the OCaml type-checker by default, but expand aliases
or type definitions if the user asks for the type at the same position
again -- this is a simple mechanism for ``how deep an answer do you
want?'' tuning. For example, asking for the type of \code{None} will
first return \code{'a option}, then give the type definition:
\code{type 'a option = None | Some of 'a}

Finally, Merlin provides a two-argument query to check the type that
a given program text would have at a given location -- even if it is
not in fact present at that location in the buffer. This is a building
block to provide user-facing features to type-check various
expressions in a local typing environment, which is important in
languages with complex scoping structures.

\subsection{\maybecap{Autocompletion}}

Autocompletion suffers from the same difficulties on the overlapping
namespace and rich name structure as type-at-point queries
(\Cref{subsec:type-at-point}).

One interesting heuristic that Merlin uses for completion is, when
there is currently nothing under the cursor, to insert a dummy/empty
identifier. For example, consider the ill-typed partial program
\texttt{(1 + \cursor)}, where $\cursor$ denotes an (empty) cursor
position; inserting a dummy identifier will allow the type-checker to
succeed and provide relevant type information for completion. The
case \texttt{(f \cursor)} is even more important, as inserting an
identifier creates an application node, providing type information
from \code{f}'s argument type.

\subsection{\maybecap{Go to definition}}

The meaning of the ``go to definition'' action is harder to match to
user's expectations in presence of a rich module system. Users may
expect the ``definition'' to follow inclusion of modules,
equalities/aliases of modules, equalities/aliases of values,
etc. Consider the following functor definitions:

\lstinputlisting{code/features_locate_1.ml}
\lstinputlisting{code/features_locate_2.ml}

Currently, resolving \code{Mi.f} for any of the modules \code{Mi}
defined above will bring the user to the definition of \code{Mi} as
a functor application. Some libraries in the OCaml ecosystem, such as
Jane Street's Core library, make a heavy use of functors and module
inclusion, and their user have reported the expectation that the
functor definition would be returned, or even, in all cases except
\code{Twice} where this does not make sense, that the definition site
of \code{A.f} be returned. This seems to call for a new form of ``how
deep an answer do you want?'' interface behavior.

\subsection{\maybecap{Destruct}}

The \texttt{destruct} feature of Merlin generates pattern-matching
clauses to improve pattern coverage. If called on an expression, it will
generate a complete cover of depth one:
\begin{itemize}
    \item for a sum type: one clause par variant constructor,
      with a wildcard pattern for each constructor argument
    \item for a record: a single record pattern, binding each record
      field to a variable of the same name
\end{itemize}

When called on a pattern, if the match is partial it will add as many
clauses as required to make it total. Alternatively, if the pattern
under the cursor is a catch-all pattern (wildcard \code{_}
or variable), then it replaces it by the various cases that it could
have matched.

This is surprisingly easy to do in Merlin thanks to the interface of
the exhaustive-match checker in the OCaml compiler, which returns the
depth-one examples of non-matched patterns in case exhaustivity
checking fails. The moral of that story would be that a good compiler
interface (for end-users) can often be usefully relied on by language
servers. (Merlin does not parse a warning destined to end-users, it
calls an internal function that generates the counter-example in
AST form.)

\subsection{\maybecap{Hoogle}}

Merlin implements a limited \Hoogle{}-like functionality, namely the
search for a type in the compilation environment which contains
certain subterms at a given polarity (negative or positive occurrences
in the type). For example, passing the search string \code{-ltac_trace
  +Pp.std_ppcmds} within a Coq development environment will give
a single result
\lstinputlisting{code/features_hoogle_1.ml}
where the pretty-printer of tactic traces is found even though the
output type is wrapped under (covariant) datatypes.

This is convenient and not particularly hard to implement with the
typechecker codebase at hand, but on the other hand providing a user
interface for this feature is difficult. Most editors do not offer
a convenient interface to type such queries, and the feature is much
less discoverable than other Merlin features.
\endgroup}

\Short{}{\SectionFeatures}

\Short{}{
\section{\maybecap{Difficulties}}
\label{sec:difficulties}
}

\newcommand{\SectionDifficulties}{\begingroup
\subsection{\maybecap{Merlin at scale}}
\label{subsec:at-scale}

Merlin has been adopted by virtually all programmers who
discovered OCaml in the past few years, no matter their programming editor
of choice -- only some venerable ancients still resist the idea of
having to tweak their Emacs configuration one more time.

In particular, Merlin is used at Jane Street, a company with a large
OCaml codebase whose programming idioms involve sophisticated module
constructs, and has financially supported Merlin's development from
its early stages. Their use of Merlin hit various performance
bottlenecks, some of which are common in software tooling use on large
codebases (Git, Mercurial had to be adapted for entreprise monorepos),
while others stem from the interactive use of a compiler codebase not
designed for interactivity.

\subsubsection{\maybecap{Short paths}}
\label{subsubsec:short-paths}

One such case is the use of the \code{-short-paths} compiler option.
This option, added in OCaml 4.01.0 (September 2013), instructs the
compiler to try hard to pretty-print, among the many different
equivalent representations of a type (modulo aliasing, expansion of
parametrized types, etc.), the shortest one. The problem with the
initial implementation of this option is two-fold:
\begin{itemize}
\item It wasn't conceived with the interactive use of merlin in mind
  (the compiler typically needs to print types for error messages, and
  it only ever prints one error per run). A key data structure in the
  algorithm is a map that records path compression choices, and this
  map was constructed eagerly, in a way that was not amenable to
  incrementality.
\item There were various hidden performance bottlenecks that were not
  apparent on small codebases.
\end{itemize}

As a result, the use of this (optional) feature in Merlin resulted in
major slowdowns: printing a type, which would previously take a few
hundred milliseconds, would now take several seconds on average. The
Merlin authors added some caching and made some effort to improve
incrementality, but this did not suffice to make the option
pleasant. Eventually Leo White, from Jane Street, did a complete
rewrite of this option. The new implementation, built with Merlin's
use case in mind, gives much better performances. After having been
beta-tested in Merlin for some months, this new implementation is
about to be proposed for inclusion in the compiler itself.

\begin{figure}[!ht]
  \centering
  \includegraphics[width=6cm]{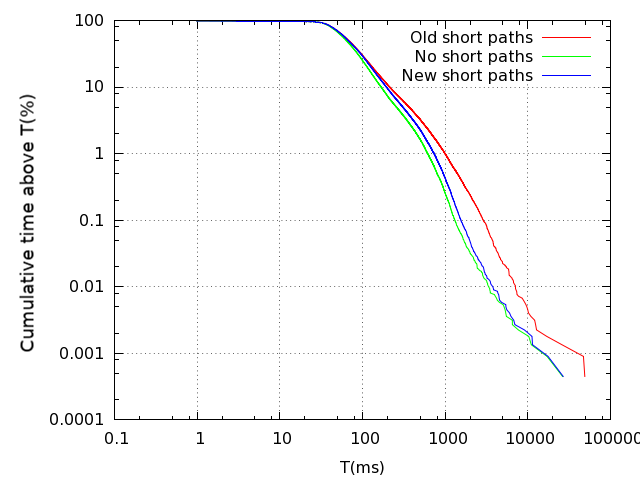}
  \includegraphics[width=6cm]{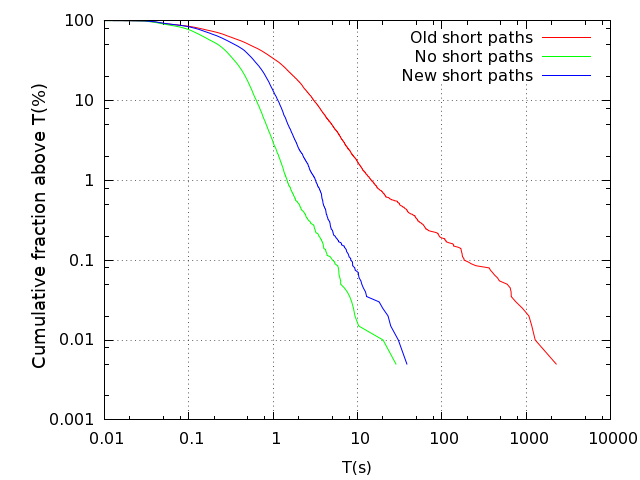}
  \caption{Cumulative time distribution of type printing.\\
    Left: \texttt{type-at-point} queries in Merlin.\\
    Right: printing the interface of Jane Street's compilation units.}
  \label{fig:short-path-graph}
\end{figure}

Leo White kindly provided us with a graph to demonstrate the
performance impact of a reimplementation. The first graph was produced
by generating many \texttt{type-at-point} queries in
random places of the Jane Street codebase. It is included in
Figure~\ref{fig:short-path-graph}. One can observe, for example, that
the number of queries taking more than 1s to complete is halved by the
new implementation, and the number of queries taking more than 3s
moved from 1/1000 to 1/10,000. The second graph measured printing the
typed interface of Jane Street's compilation units, and the difference
are more dramatic.

Our take-away from this story is that language servers can reveal or
magnify performance bottlenecks in a language frontend, and serve as
experimentation grounds before ideas are upstreamed in the language
implementation.

\subsection{\maybecap{Reason}}

\Reason{} is an alternative syntax for OCaml that started as
a Facebook project. It brings a syntax close to Javascript while
leaving the semantics of OCaml untouched. A lot of efforts are poured
into integrating seamlessly with the Javascript ecosystem. The aim is
to provide a smooth experience to Javascript developers, and to allow
gradual or partial transitions of Javascript codebase to an ML style.

Since the semantics of OCaml is retained, Merlin could be made to work
with Reason codebases without too much work. In fact, Merlin played an
important role in kick starting Reason by offering pleasant tooling at
a low cost from the early days.

To enable support for alternative syntaxes, the abstraction of a
\emph{reader} was introduced in Merlin. A reader is a component that can
customize the build environment, the lexer or the parser, and a code
pretty-printer -- to show, for example, error messages in the user's
input syntax. It takes the form of an helper process that is managed
by Merlin and exposes a few commands that can redefine some of
Merlin's behavior. A small library, \code{merlin-extend}, serves as an
interface between Merlin and Reason (or any other alternative
OCaml frontend): providing an alternative frontend is just a matter of
instantiating a functor and generating a binary, to be found in
Merlin's PATH. Currently the selection of a reader is made based on
the source file extension: \code{.ml}, \code{.mli} for OCaml,
\code{.re}, \code{.rei} for Reason.

We believe that the ease with which Merlin was adapted to support
a second language frontend -- while the frontend is arguably the
central piece of a language server -- validates its design choices.

\subsection{\maybecap{Communicating with editors}}

Here come some bad news. First, scripting text editors (which is
required to build an editor-mode to talk to Merlin) remains a painful
affair -- the intensity varies per editor. Second, the open source
editors whose admittedly arcane keybindings your grandparents proudly
taught you are the worst offenders, and the Javascript-implemented
monstrosities of our decadent age are sensibly better.

There are many things to be said about Emacs or Vim in the context of
language-server mode implementations. Here are some of the stories we
heard from scarred developers -- the following is only hearsay:
\begin{itemize}
\item Their concurrency and process handling APIs are terrible.
\item They are full of modes for each user-level
  interface aspects (indentation, syntax coloring, interactive error
  feedback, auto-completion etc.) that do not talk to each
  other. Every user has their own point in the combinatorial space of
  mode combinations, but most users don't actually know which point it
  is or how to configure it.
\item Those modes will implement extremely ad-hoc strategies
  that have proved to do a decent job for C and Python in
  the past. Auto-completion modes will use incorrect caching
  strategies that you cannot easily work around. Interactive error
  checkers such as \code{flymake} will offer a public API
  expecting a shell command, but no script-language function. One of
  the popular modes to display/highlight source position to the user
  only accepts \emph{source code regular expressions} to describe
  positions!
\end{itemize}
(In contrast, they described Visual Studio Code as offering a much
better scripting experience.)

\label{subsec:LSP}

There are two strategies to avoid editor-mode burnout. One is to
let users who really love their text editor write the mode for it --
Merlin has had some success in this area. The other is to reuse the
Language Server Protocol (LSP), an open exchange format for
interactions between program editors and language servers proposed in
2015, initially for Visual Studio Code.

The LSP is a nice idea, and is quickly gaining adoption among editors
and language servers. It is not yet the premier interaction format for
Merlin, but
a \href{https://github.com/freebroccolo/ocaml-language-server}{LSP
  frontend} has been built on top of Merlin and may get used more and
more in the future.
\endgroup}

\Short{}{\SectionDifficulties}

\Short{}{
\section{\maybecap{Thoughts on language tooling}}
\label{sec:thoughts}
}

\newcommand{\SectionThoughts}{\begingroup
\subsection{\maybecap{No spec, no tests}}

The Merlin authors have found it very difficult to write useful
testsuites for Merlin. The specification of user-facing features is
generally very unclear, and it evolves along with the tools. Merlin
remains debugged mostly by using it on real projects. The developers
are in the process of building a ``feature map'' document delimiting
a scope, and a set of well-identified features or concepts which have
a clear testing strategy.

\subsection{\maybecap{The sufficiently smart language server}}

Users of interactive modes tend to have higher, weirder expectations
on the output they get from their tools, compared to batch
compilers. People know that compilers sometimes emit unintuitive or
surprising messages, but the experience of a language server,
being more direct, seems to result in users reacting more strongly to
expectation/behavior mismatches.

This can get amplified by recovery strategies which produce completely
useless results in some bad cases. It is tempting to look for
a classification of which feedback outputs are likely to be unhelpful,
and revert to safer heuristics (and less output) in that case; but
there is no clear specification of such a classification.

\subsection{\maybecap{Fragility of the OCaml ecosystem}}

The relative fragility of the OCaml tooling ecosystem, in particular
regarding to build specifications, is a difficulty for Merlin. The
OCaml compiler provides a \code{cc}-like interface, building one
single file in a given compilation environment, and several competing
build systems are used within the community to turn that into a more
unified ``build my project'' experience.

Users do not wish to duplicate their build configuration twice, once
for the build system and another for Merlin. Merlin will try to
identify the root of the software project and consider all compiled
interface files in the filesystem subtree, which works well for simple
projects, but for example it does not currently provide a way to pass
per-file build options or configuration environments. The recent
\code{dune} build system knows how to generate Merlin configuration
files, but these limitations in Merlin's project description language
remain problematic.

There seems to be no easy solution out of this; some software
ecosystems are more centralized and have a single blessed system for
project description and build description, which makes tooling
easier. Otherwise, we can only hope that the various project
description and build systems can generate a description of each
file's compilation environment in a common format -- there is
discussion within the OCaml community of adopting Clang's
\href{https://clang.llvm.org/docs/JSONCompilationDatabase.html}{JSON
  Compilation Database} for its purpose.

This is a problem shared with other language servers. The LSP protocol
which we mentioned in \Cref{subsec:LSP} has been slowly
evolving to add a notion of \emph{project}: instead of just having
a separate session per file buffer, files would be understood within
some build context.

\subsection{\maybecap{Language constructs for documentation and feedback?}}

We believe that there is an opportunity for language design efforts to
include considerations regarding editor-inspired notions, such as
tracking where a name provided by a module is defined and
documented.

Consider sealing a module with a signature for example,
\code{module N = (M : S)}.
A language server could track, when asked questions about \code{N.f},
how this module was defined, and know to follow this definition to
inspect \code{M}, if asked about the definition/implementation of
\code{f}, or to inspect \code{S}, if asked about the declaration or
documentation of \code{f}.

However, it could be the responsibility of the type system to carry
around, in the signature of \code{N}, not just the static types, but
also documentation and other interface metadata coming from
\code{S}. (An early work in this direction is the interaction between
documentation and the module system in the Scribble documentation
system for Racket~\citep*{scribble}.)
\endgroup}

\Short{}{\SectionThoughts}

\section{\maybecap{Conclusion}}

\subsection{\maybecap{Future work}}

\subsubsection{\maybecap{Memoized LR reductions}}

It is possible to memoize the effect of parsing a sequence of tokens
from a given LR automaton state, compressing a series of shift/reduce
actions into a single effect on the parsing stack, the addition of
a semantic value. Implementing this for Merlin is planned; it would
allow a form of incrementality that can reuse work not only on valid
(or recovered) prefixes on the input, but also share work across
different parts of the buffer, and instantly re-parse input that was
intermittently placed after an unrecoverable error.

\subsubsection{\maybecap{Menhir development kit}}

Merlin changed Menhir to support adding arbitrary annotations to
a grammar, and processing those annotations externally -- a new Menhir
mode provides access to serialized forms of both the annotated grammar
and the compiled parsing automaton. This is growing into a more
general ``Menhir SDK'' library, allowing various previously-unplanned
uses of the grammar and automaton. For example, a work in progress is an
algorithm to pretty-print parsing derivations on the concrete syntax,
which is useful for teaching and grammar-engineering purposes.

Menhir supports parametric rules, abstracting some rules over a symbol
or sequence of symbols, and inlining applications to specific
parameters at compile-time. Their presence creates a distance between
the user grammar and the grammar exposed by the introspection API,
which is the post-inlining grammar. There is work ongoing on
``un-inlining'' approaches to recover applications of parametrized
rules, but higher-order parameters make this challenging or impossible
in the general case.

\subsection{\maybecap{Related Work}}

\subsubsection{\maybecap{Tim Wagner's work on Incremental Parsing}}

An interesting body of work on incremental parsing is Tim Wagner's PhD
thesis work in the
1990s~\citep*{wagner-incremental-parsing,wagner-error-recovery,wagner-phd}. The
Merlin authors were not aware of this work as they designed their
incremental frontend, so it is interesting to compare their
approaches.

Wagner's work studies lexer and parser generators as used in
imperative and object-oriented programming languages; general
differences with the Merlin work arise from this difference in
paradigm. A lexer or parser generator for OCaml will typically
generate a nest of tail-recursive functions; on the other hand,
a C implementation would typically work with a single execution loop,
mutating a bit of state that represents the control (the next
automaton state to consider). In other words, lexers and parsers in
imperative languages tend to have their control flow reified in data and
define their semantic actions as manipulating mutable values. Lexers
and parser in functional language tend to have an abstract control
flow, but return immutable values. Incremental lexing and parsing
requires the best of both world: reified control flow, and pure
actions returning immutable values. Wagner had to take care of
isolating actions and preserving meaningful notions of physical
identity to rely on; Merlin can just compare pure values, but had to
add continuation-passing or control-inversion mechanisms.

Wager also points out that, for proper incrementality support,
lookahead must be carefully handled: a lexer or parser decision may
depend on what follows next in the buffer, and be invalidated by
buffer changes. LR parsers typically use a single lookahead token, but
lexers may use arbitrary lookahead; Wagner keeps track of the dynamic
lookahead for incremental lexing. This is not implemented in Merlin,
as the OCaml lexer does not require arbitrary lookahead, but it may be
important for adaptation to other languages.

Finally, Wagner established some complexity results, characterizing
notions of optimal algorithms for incremental processing. In
particular, parsing spines may sometimes have to be rebalanced to
preserve logarithmic access costs and avoid linear worst-cases. These
results typically do not include the cost of producing the AST
structure, or require changes to this structure -- rebalancing
a partial list as a tree for logarithmic access is pointless if the
semantic action then immediately builds a linear list to represent
this AST.

An implementation inspired by Tim Wagner's work is the
\href{https://github.com/tree-sitter/tree-sitter/}{Tree-Sitter}
project developed by Max Brunsfeld at Github. One important difference is
that Tree-Sitter uses a GLR parser algorithm. This project seems
focused on providing reasonable (but possibly approximate) parsers for
a large number of languages, for integration within the Atom code
editor.

Tree-sitter also implements a ``contextual scanner'', a lexer that is
indexed over the parser state. This has been done manually in some
OCaml projects, but it could be added to Menhir as a general feature.

Another, closer implementation of Tim Wagner's ideas can be found in
the work on the Eco editor that we discuss later in this section. The
Eco project did not implement AST rebalancing, and the authors report
that this does not seem to create performance issues in practice.

\subsubsection{\maybecap{Yacc error recovery}}

It is interesting to compare Merlin's approach to syntax error
recovery with Yacc's error recovery using the \code{ERROR}
tokens. Yacc lets users integrate error tokens in any rule, to
print a user-written error message or to provide a default action for
recovery. If a parsing error is encountered, it will typically not be
exactly in the place of an error token in an error-handling
rule, so Yacc has an algorithm to perform ``error reductions''
until it finds an enclosing error-handling rule.

Starting from the same state, the strategies can be summarized as
"Yacc drops information that separates it from an error handling
rule", while "Merlin invents information that separates it from
a possible recovery".

The benefits of Merlin's strategy are significant for interactive use:
while inventing information could mean generating garbage, the process
can actually be guided by the user. Preserving information means that the
last tokens typed by the user are considered during parsing and further
analyses.

For instance, in \code{let f x : int =}, the prefix is correct: Merlin
knows the context should be completed by an expression of type int and
that a variable \code{x} is in scope. With Yacc, this property only
holds if there is an error-handling rule covering exactly this
case. It is difficult to add enough error-handling rules: there will
need to be many, and they should not conflict.

Merlin is also more flexible in the handling of ambiguities: while
Yacc rejects ambiguous error rules (causing shift/reduce or
reduce/reduce conflicts), Merlin allows ambiguities and uses its cost
model for token synthesis to direct the non-deterministic search.

\subsubsection{\maybecap{Eco}}

The Eco editor~\citep*{eco} supports language composition (mixing of
programming languages within the same editor file), and relies on
incremental parsing to provide syntax-aware editing. Eco bridges the
gap between syntax-directed editing and textual editing by using as
its main source representation a concrete syntax tree
(incrementally constructed) structured to retain all textual
information, in particular whitespace and comments.

Merlin uses a less principled approach, working with parse trees and
ASTs, and relying on source positions stored in the AST to map AST
nodes to source text fragments.

\subsubsection{\maybecap{Editor modes}}

Many IDEs provide language-aware support, generally for mainstream
languages with a large user base: C/C++, Java, etc. An early example
is the Delphi IDE for Pascal. Those systems offer a very fine-grained
integration of the language -- Delphi directly hooks into its own
Pascal compiler. For people who are not programming language
researchers, an IDE may be their main way to familiarize themselves
with details of the language features (scope resolution, type inference...).

On the other hand, language-aware support for functional languages
with a rich type system has a more checkered history. For example, the
Scala community went through many iterations of its incremental
type-checker. IntelliJ for Scala, for example, would use a type
checker with slightly different behavior from the Scala compiler,
resulting in inconsistencies that users would find jarring.

An interesting design space is the editor modes for proof assistants,
that brought several innovations such as typed holes. In a proof
assistant, the editor experience is central, as proofs are rarely run;
in some sense, the frontend provides the user experience.

\begin{acks}
  The main implementors of Merlin are Fr{\'e}d{\'e}ric Bour and Thomas
  Refis, two vim users -- with financial support from Jane
  Street. Simon Castellan added support for Emacs, Luc Rocher for
  Sublime Text, Hackwaly for Visual Studio Code, Facebook for Atom,
  Rapha{\"e}l Proust for Acme. Darrin Morrison implemented the Language
  Server Protocol support, David Allsopp ported the tool on Windows,
  and Gemma Gordon helped with project management.

  The mechanisms for parser-agnostic syntactic recovery was designed
  and implemented in Menhir in collaboration with Philipp Haselwarter,
  during a Ljubljana visit funded by Andrej Bauer.

  We also wish to thanks the anonymous reviewers, and Lukas Diekmann,
  for their detailed comments.
\end{acks}

\Short{\nocite{*}}{}
\bibliography{merlin}

\begin{version}{\Appendices}
\appendix

\section{\maybecap{Adapting the rest of the frontend}}
\label{sec:frontendaux}

The changes described in \Cref{sec:frontend} added two new properties to
the OCaml frontend: incrementality for performance and error recovery for
robustness and availability.

Although each step of the pipeline is affected, only the most challenging parts
were explained. In this section, we detail the changes to the remaining ones.

\subsection{\maybecap{Incremental lexing: adapting OCamllex}}
\label{subapp:incremental-lexing}

\SectionIncrementalLexing

\subsection{\maybecap{Incremental typing, sort of}}
\label{subapp:incremental-typing}

\SectionIncrementalTyping

\subsection{\maybecap{No need for lexing recovery}}
\label{subapp:lexing-recovery}

\SectionRecoveryLexing

\subsection{\maybecap{Typing recovery}}
\label{subapp:typing-recovery}

\SectionRecoveryTyping

\section{\maybecap{Features of interest}}
\label{app:features}

\SectionFeatures

\section{\maybecap{Difficulties}}
\label{app:difficulties}

\SectionDifficulties

\section{\maybecap{Thoughts on language tooling}}
\label{app:thoughts}

\SectionThoughts

\end{version}

\end{document}